\documentclass[journal=jacsat,manuscript=article]{achemso}
\setkeys{acs}{articletitle = true}
\usepackage[version=3]{mhchem} % Formula subscripts using \ce{}

\usepackage{amsmath}  
\usepackage{amsfonts} 
\usepackage{subfig}
\usepackage{graphicx} 
\usepackage{booktabs}
\usepackage{threeparttable}
\usepackage[font=footnotesize]{caption}
\usepackage{verbatim}
\usepackage{xcolor}
\usepackage{soul}

\author{James Shee}
\email{js4564@columbia.edu}
\affiliation{Department of Chemistry, Columbia University, 3000 Broadway, New York, NY, 10027}
\author{Evan J. Arthur}
\affiliation{Schrödinger Inc., 120 West 45th Street, New York, NY 1003}
\author{Shiwei Zhang}
\affiliation{Center for Computational Quantum Physics, Flatiron Institute, 162 5th Avenue, New York, NY 10010}
\alsoaffiliation{Department of Physics, College of William and Mary, Williamsburg, VA 23187}
\author{David R. Reichman}
\author{Richard A. Friesner}
\affiliation{Department of Chemistry, Columbia University, 3000 Broadway, New York, NY, 10027}

\title{Singlet-Triplet Energy Gaps of Organic Biradicals and Polyacenes with Auxiliary-Field Quantum Monte Carlo}

\begin{document}

%\begin{tocentry}
%Some journals require a graphical entry for the Table of Contents.
%This should be laid out ``print ready'' so that the sizing of the
%Inside the \texttt{tocentry} environment, the font used is Helvetica
%8\,pt, as required by \emph{Journal of the American Chemical
%The surrounding frame is 9\,cm by 3.5\,cm, which is the maximum
%permitted for  \emph{Journal of the American Chemical Society}
%graphical table of content entries. The box will not resize if the
%content is too big: instead it will overflow the edge of the box.
%
%This box and the associated title will always be printed on a
%separate page at the end of the document.
%
%\end{tocentry}

\begin{abstract}
The energy gap between the lowest-lying singlet and triplet states is an important quantity in chemical photocatalysis, with relevant applications ranging from triplet fusion in optical upconversion to the design of organic light-emitting devices.  The \emph{ab initio} prediction of singlet-triplet (ST) gaps is challenging due to the potentially biradical nature of the involved states, combined with the potentially large size of relevant molecules.  In this work, we show that phaseless auxiliary-field quantum Monte Carlo (ph-AFQMC) can accurately predict ST gaps for chemical systems with singlet states of highly biradical nature,  including a set of 13 small molecules and the ortho-, meta-, and para- isomers of benzyne.  With respect to gas-phase experiments, ph-AFQMC using CASSCF trial wavefunctions achieves a mean averaged error of $\sim$1 kcal/mol.  Furthermore, we find that in the context of a spin-projection technique, ph-AFQMC using unrestricted single-determinant trial wavefunctions, which can be readily obtained for even very large systems, produces equivalently high accuracy.  We proceed to show that this scalable methodology is capable of yielding accurate ST gaps for all linear polyacenes for which experimental measurements exist, i.e. naphthalene, anthracene, tetracene, and pentacene.  Our results suggest a protocol for selecting either unrestricted Hartree-Fock or Kohn-Sham orbitals for the single-determinant trial wavefunction, based on the extent of spin-contamination.  These findings pave the way for future investigations of specific photochemical processes involving large molecules with substantial biradical character.
\end{abstract}


%%%%%%%%%%%%%%%%%%%%%%
\section{Introduction}
%%%%%%%%%%%%%%%%%%%%%%

The energy gap separating the lowest-lying singlet and triplet states of a molecule is an important property relevant to many chemical processes.  For example, light absorption by chlorophyll in Photosystem II can produce triplet states which in turn react with triplet oxygen to produce short-lived and highly reactive singlet oxygen.\cite{krieger2005singlet}  Additionally, the relative energetics of first-excited singlet and triplet states in dopants utilized in organic light-emitting diodes governs the efficiency of such devices, and is a useful parameter for the design of light-emitting electronics\cite{kwak2016silico}.  In the field of photocatalysis the singlet-triplet (ST) gap is directly relevant to a variety of redox reactions.\cite{PhotoCatReview}  In addition the ST gap is a quantity of crucial importance for determining the energetic feasibility of the optical processes known as singlet fission\cite{SingletFissionChemRev} and upconversion\cite{UpconversionReview}.  In the former process, a single photon produces a high energy singlet state which eventually splits into two triplet excitons; in the latter, two triplet excitons annihilate or fuse to form a high energy emissive singlet.  In certain cases the ST gap can be probed under cryogenic conditions via phosphorescence measurements, however for a large number of relevant molecules direct measurement is not possible. 

The electronic structures relevant to these types of applications can be complicated by the presence of biradical character in one or more of the involved spin-states.  Biradicals are molecules in which two valence electrons can occupy two degenerate but spatially distinct molecular orbitals\cite{Salem} (the species are referred to as biradicaloids if these two orbitals are nearly-degenerate, but we do not make this distinction here).  
%In general, the relative ordering of low-lying spin-states is complicated and depends, in simple models, on factors such as the spatial overlap, coulomb and exchange energies, and the difference in orbital energies.\cite{goldberg1983effects,Abe_Review,Salem}  
The many possible electronic configurations give rise to their capacity to exhibit remarkably specific chemical reactivity.\cite{Abe_Review,moss2004reactive,carbeneRev,RovisCarbeneRev,kirmse2013carbene}. Singlet states can be characterized as either closed-shell or open-shell.  In the former, one of the two valence states is doubly occupied, which is typically the case, e.g., in carbenes.  These species can simultaneously display Lewis base and Lewis acid character, and thereby undergo concerted addition reactions, e.g. with alkenes, to produce stereospecific products.  Open-shell singlet states\cite{gopalakrishna2018open} are characterized by single-occupancy of each of the two valence states, as in all triplet states due to Pauli exclusion.  Such open-shell molecules yield products of mixed stereochemistry.

The electronic structure community has witnessed the development of promising theoretical methods for computing the ST gap in biradical molecules.  This is a challenging problem for traditional single-reference \emph{ab initio} methods, e.g. Hartree-Fock (HF) and Density Functional Theory (DFT), as a minimal quantum-mechanical description of the wavefunctions corresponding to all singlet states and one triplet state ($M_s = 0$) in the two electron two orbital model is necessarily a superposition of two electronic configurations,\cite{KrylovSpinFlip,yang2015singlet} thus requiring more than one Slater determinant.  Even for simple chemical species such as NH and O$_2$, which have triplet ground states, the singlet spin-configuration is notoriously difficult to describe.  Accurate predictions are further complicated by the requirement that static and dynamic electron correlation be well balanced, with spin-states of both multiplicities treated on equal theoretical footing. Biradical systems have been studied with DFT\cite{bendikov2004oligoacenes,squires1998electronic} and fractional spin variants,\cite{ess2010singlet,VariationalFSDFT} Generalized Valence Bond theory,\cite{carter1987electron,carter1987new,carter1988correlation,irikura1992singlet}
spin-flip (SF) methods,\cite{KrylovSpinFlip,SFTDDFTmhg,KrylovEOMSF,bernard2012general} 
%(SF-CIS and SF-TDDFT can achieve $O(r^4)$ scaling)
Complete Active Space Self-Consistent Field (CASSCF) with 2nd order perturbation theory,\cite{stoneburner2017systematic} 
multiconfiguration pair-density functional theory (pDFT),\cite{bao2016correlated,wilbraham2017multiconfiguration,GhoshST,stoneburner2018mc,SharmaAcene}
Coupled Cluster (CC) methods,\cite{FPAacenesFirst,FPAacenes,wloch2007extension} 
doubly electron-attached equation-of-motion CC theory, \cite{shen2013doubly,ajala2017economical}
spin-extended Configuration Interaction (CI) with singles and doubles, \cite{tsuchimochi2017bridging}
incremental Full CI,\cite{zimmerman2017singlet}
Difference Dedicated CI,\cite{garcia1996singlet,cabrero1999singlet,calzado2011analysis}
the particle-particle random phase approximation (pp-RPA)\cite{yang2015singlet,YangPNAS} 
%($r^4$, $r=max(N_{occ}, N_{virt})$)
the Density Matrix Renormalization Group (DMRG),\cite{hachmann2007radical}
Yamaguchi spin projection\cite{YAMAGUCHI1988537} and its recent combination with orbital-optimized MP2\cite{lee2019two}. 

In this work we use phaseless auxiliary-field Quantum Monte Carlo (ph-AFQMC)\cite{zhang1995constrained,zhang2003quantum} to compute ST gaps.  While imaginary-time projection is most frequently used to yield ground-state properties of a system, the formalism has also been used to accurately compute low-lying excited states of materials\cite{ma2013excited}, molecular diatomics\cite{purwanto2009excited} and dipole-bound species.\cite{DipoleBoundAnions}  These calculations rely on the fact that eigenfunctions of the Hamiltonian are orthogonal.  In practice, when the exact eigenfunctions are unknown, ph-AFQMC calculations use a so-called ``trial wavefunction'' to project out orthogonal components that may be sampled along the imaginary-time random walks.  For example, for the molecules with triplet ground-states relevant to this work, the energy of the lowest-lying excited singlet state can be sampled using a trial wavefunction with $\langle S^2 \rangle = 0 $, due to its near-orthogonality to the true triplet ground-state.  In the limit of an exact trial wavefunction, this symmetry-constraining approach would be exact.  

In what follows we will show that an unrestricted single-determinant trial  wavefunction is capable of accurately describing multi-reference biradical species.  This is a significant result because the computational cost of CI and CASSCF calculations, despite many recent advances,\cite{ZgidDMRGSCF,CASgpu,HCISCF,parallelMCSCF} scales exponentially with system size and thus such methods are infeasible as trial wavefunctions for ph-AFQMC.  As an example, the number of $\pi$ electrons in the polyacene series is 4$n$+2 ($n$=1,2,... for benzene, naphthalene,...), which typically must be included in the active space for accurate MCSCF-based predictions.  Typical CI solvers can handle up to $\sim16$ active electrons and orbitals, which would be insufficient for $n > 3$.  

We view the ability of an electronic structure theory to accurately describe biradicals as a prerequisite for future studies of large, typically conjugated systems that catalyze photochemical processes such as upconversion.  After showing that a spin-projected approach to ph-AFQMC\cite{purwanto2008eliminating} with unrestricted single-determinant trial wavefunctions produces accurate results for a set of strongly biradical small molecules and three benzyne isomers, we then illustrate the scalability of our approach by taking a first step toward relevant and relatively large photocatalytic molecules, namely the polyacene series including naphthalene, anthracene, tetracene, and pentacene.  These molecules are well-studied both experimentally (with measurements available in the literature for $n$=1-5, though not beyond) and theoretically, as they have myriad  applications in organic electronics (see references in the first paragraph of Ref. \citenum{YangPNAS}).  While biradical and polyradical character is predicted to be responsible for the instability of hexacene and longer acenes,\cite{hachmann2007radical,ClarSextetRev} we focus on the $n$=2-5 molecules since they are representative of the majority of molecules in our target class of photocatalysts.\cite{PhotoCatReview}  In fact, $n$=3,4 are known to perform upconversion,\cite{UpconversionReview} and $n$=3-5 are known singlet-fission catalysts.\cite{SingletFissionChemRev}

This work is organized as follows.  Details of our computational approach are described in Sec. II.  In Sec. III we compute ST gaps for a set of 13 small organic molecules which have singlet states of highly biradical nature, and compare with experimental measurements.  Next we examine ortho- meta- and para- isomers of benzyne, and show that very high accuracy can be obtained with both CASSCF and single-determinant trial wavefunctions using a basis set of moderate size.  Having shown that a single-determinant trial wavefunction combined with a spin projection technique is capable of accurately describing multi-reference biradical species, we proceed to compute ST gaps with ph-AFQMC for the increasingly large (but not necessarily biradical) systems naphthalene, anthracene, tetracene, and pentacene, and compare our results with state-of-the-art electronic structure theories and experimental measurements.  
 
%%%%%%%%%%%%%%%%%%%%%%%%%%%%%%%
\section{Computational Details} 
%%%%%%%%%%%%%%%%%%%%%%%%%%%%%%%

The ph-AFQMC methodology is reviewed in Refs. \citenum{motta2018ab} and \citenum{zhang2018ab}.
All calculations utilize an imaginary time step of $\Delta\tau = 0.005 Ha^{-1}$.  Walker orthonormalization, population control, and local energy measurements are carried out every 2, 20, and 20 steps, respectively.   We utilize a modified Cholesky decomposition of the electron repulsion integrals,\cite{purwanto2011assessing} with cutoffs of $10^{-5}$ for the small molecule biradicals, and $10^{-4}$ for polyacenes, $n$=2-4.  For pentacene ($n$=5) we use density fitting with the Weigend Coulomb-fitting basis set\cite{weigend2006accurate} to reduce the memory requirements of the calculation, while preserving high accuracy in energy differences\cite{shee2018gpu}.  

All calculations utilize single-precision floating point arithmetic.  For ionization and bond-dissociation energies of transition metal atoms and diatomics, respectively, this yielded very high accuracy while reducing the computational cost compared to double-precision calculations\cite{shee2018gpu,shee2019}.  
\begin{comment}
For pentacene, the largest molecule considered in this work, the difference in the single- and double- precision energies using the STO-3G basis is 0.0018(21)$Ha$ for the singlet state, and 0.0017(18)$Ha$ for the triplet state.  The differences in the measured energies due to the use of single-precision are smaller than the statistical error bars.
\end{comment}
For pentacene, the largest molecule considered in this work, we verified with separate calculations in the STO-3G basis that single- and double-precision calculations gave statistically indistinguishable results.

For the small molecule biradicals, we use the aug-cc-pV$x$Z basis sets,\cite{kendall1992electron} with $x$=T,Q.  Unrestricted HF (UHF), restricted HF (RHF) and its open-shell variant (ROHF), unrestricted Kohn-Sham DFT with the B3LYP functional (UB3LYP), and CASSCF trial wavefunctions are obtained using PySCF\cite{sun2018pyscf}.  ST gaps are extrapolated to the complete basis set (CBS) limit using exponential and $1/x^3$ forms for the mean-field and correlation energies, respectively, as detailed in, e.g., Ref. \citenum{shee2018gpu}.

For the benzyne isomers and polyacenes we report ST gaps in the cc-pVTZ basis,\cite{dunning1989gaussian} primarily for computational expedience, since pentacene has nearly 900 basis functions.  For hydrocarbon systems, full CBS extrapolation typically alters the triple-zeta results by $\sim$1 kcal/mol or less.  This empirical finding is consistent with previous studies of polyacenes using other wavefunction methods which show little basis set dependence.\cite{FPAacenesFirst,Evangelista_sCI,YangPNAS}  %Indeed pp-RPA in just a double-zeta basis still compares very well ($<$ 2 kcal/mol) with experimental measurements.  

%(if using Corr Samp data, precedence for using MP2 for geometry relaxation energy on biradical set\cite{garcia1996singlet})

Our ph-AFQMC calculations utilize the spin-projection technique detailed in Ref.\citenum{purwanto2008eliminating}.  The walkers are initialized with RHF for singlets and ROHF for triplets, which have $\langle S^2 \rangle$ of exactly 0 and 2, respectively.  With an appropriate form of the Hubbard-Stratonovich transformation this can ensure that the single-particle imaginary-time propagator preserves spin-symmetry despite the use of a (possibly) spin-contaminated unrestricted trial wavefunction.  

In what follows, we propose a simple empirical protocol, ``AFQMC/U", to guide the selection of an optimal unrestricted orbital set, among UHF and UB3LYP orbitals, to be used as the trial wavefunction in AFQMC.  UHF orbitals are used by default, unless $\langle S^2 \rangle_{singlet}^{UHF}$ or $\langle S^2 \rangle_{triplet}^{UHF}$ deviate by more than 10$\%$\cite{young2004computational} from their spin-pure values, i.e. 0 and 2, respectively.  For biradical singlet states, UB3LYP orbitals are utilized only when $\langle S^2 \rangle_{singlet}^{UHF} > 1.1$, since wavefunctions constructed from unrestricted Kohn-Sham orbitals are known to exhibit less spin contamination compared to UHF solutions.\cite{baker1993spin}   Biradical triplet states are \emph{not} expected to exhibit significant spin-contamination as they are typically  well-described by a single determinant (the two-determinant $M_s$ = 0 triplet state is not encountered due to the constraint on $\langle S_z \rangle$ in the quantum-chemical methods relevant here).

%%%%%%%%%%%%%%%%%%%%%%%%%%%%%%%
\section{Results and Discussion}
%%%%%%%%%%%%%%%%%%%%%%%%%%%%%%%

\subsection{Small Molecule Biradical Set}
\begin{figure}[H]
    \centering
    \includegraphics[width=16cm]{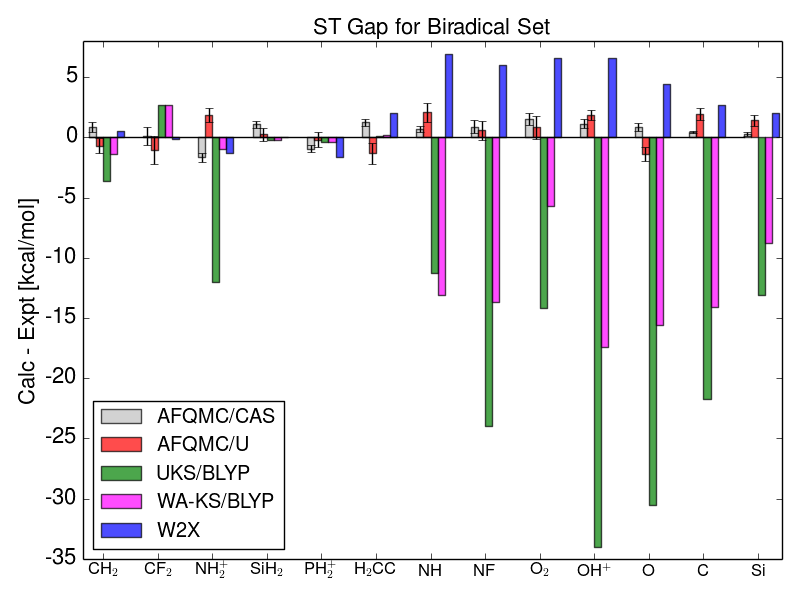}
    \caption{Deviations [kcal/mol] of various calculations from experimentally-derived reference values.  Error bars show the statistical error of the QMC measurements.} 
    \label{fig:biradicals_largeCAS}
\end{figure}

ST gaps are computed for the set of 13 small molecules with singlet states that exhibit substantial biradical character, recently examined in Ref. \citenum{bao2016correlated}. Molecular geometries for the biradical species are taken from the Supporting Information of  Ref. \citenum{bao2016correlated}, which used QCISD/MG3S for CF$_2$ and QCISD(T)/aug-cc-pVQZ for the rest, with unrestricted (restricted) HF references for triplet (singlet) multiplicities, respectively.  ph-AFQMC results utilizing CASSCF trial wavefunctions are plotted in Fig. \ref{fig:biradicals_largeCAS}, relative to experimental reference values, along with data provided in Ref. \citenum{bao2016correlated} for broken-symmetry DFT with the BLYP functional (UKS/BLYP), its spin-projected variant (WA-KS/BLYP), and a composite method which has been shown to produce comparable accuracy to CCSD(T)/CBS\cite{chan2015w2x} (W2X).  We note that we use a different reference value for NH$_2^+$, which is from a genuine experimental measurement\cite{gibson1985photoionization}, following Ref. \citenum{electronicstatesNH2}.

The ph-AFQMC ST gaps have been converged with respect to active space sizes of the CASSCF trial wavefunctions, an approach which has been shown to produce very high accuracy even for strongly correlated systems.\cite{shee2019,shee2018gpu} However the generation of such trial wavefunctions is, in practice, limited to moderate system sizes due to the procedure's exponential scaling.  In light of applications to large systems such as those found in organic electronics which we will focus on in future investigations, we explore the simplest scalable alternatives, namely UHF and RHF wavefunctions, and single determinants constructed from UB3LYP orbitals.  $\langle S^2 \rangle$ values with respect to the trial wavefunctions are shown in Table \ref{table:SpinContam}.

\begin{table}[H]
\begin{threeparttable}
\caption{$\langle S^2 \rangle$ of the singlet and triplet unrestricted solutions for the 13 small molecule biradical set.}
\begin{tabular}{l r r r r }
\hline\hline 
           &  UHF singlet & UHF triplet & UB3LYP singlet & UB3LYP triplet \\ [0.5ex] 
\hline 
CH$_2$    &  0     &  2.02 & 0.57 & 2.01  \\
CF$_2$    &  0.14  &  2.01 & 0 & 2.00  \\
NH$_2^+$  &  0.82  &  2.02 & 0.75 & 2.01  \\
SiH$_2$   &  0.43  &  2.01 & 0 & 2.00 \\
PH$_2^+$  &  0.44  &  2.01 & 0 & 2.00 \\
H$_2$CC   &  0.16  &  2.35 & 0 & 2.03 \\
NH        &  1.01  &  2.02 & 1.00 & 2.01  \\
NF        &  1.01  &  2.02 & 1.00 & 2.01 \\
O$_2$     &  1.02  &  2.05 & 1.01 & 2.01 \\
OH$^+$    &  1.01  &  2.01 & 1.00 & 2.01 \\
O         &  1.01  &  2.01 & 1.00 & 2.00 \\
C         &  1.02  &  2.01 & 1.01 & 2.00 \\
Si        &  1.05  &  2.02 & 1.01 & 2.00 \\
\hline\hline 
\end{tabular}
\label{table:SpinContam}
\end{threeparttable}
\end{table}

Table \ref{table:biradicalMAE} provides a rudimentary statistical representation of the accuracy of selected theoretical approaches with respect to experiment, comparing the mean signed error (MSE), mean absolute error (MAE), and maximum error (MaxE).  As expected, utilizing CASSCF trial wavefunctions can obtain excellent accuracy, with an MAE of less than a kcal/mol and MaxE of 1.7(3) kcal/mol for O$_2$.  For this molecule, we have tried active spaces of 12e8o, 8e12o, and 10e15o.  The latter two active spaces produce statistically indistinguishable ST gaps in the CBS limit.  Additionally, we verified that the total energies of both the singlet and triplet states in the aug-cc-pvtz basis differ by less than 1$mHa$ going from one active space to the next.  This suggests that the AFQMC/CAS result is converged with respect to active space size, and the remaining deviation from experiment is likely due to inaccuracies of the optimized geometries and/or the zero-point energies used to correct the experimental result.  

\begin{table}[H]
\begin{threeparttable}
\caption{Mean signed, absolute, and maximum errors [kcal/mol] of the theoretical methods shown in Fig. \ref{fig:biradicals_largeCAS} for the 13 small molecule biradical set. Sorted by MAE value. Parenthesis indicate statistical errors.}
\begin{tabular}{l r r r}
\hline\hline 
           &  MSE  & MAE & MaxE    \\ [0.5ex] 
\hline 
AFQMC/CAS    & 0.5(4)  & 0.9(4) & 1.7(3) \\
AFQMC/U      & 0.5(7)  & 1.2(7) & 2.1(8) \\      
W2X        & 2.7 & 3.1 & 6.9 \\
WA-KS/BLYP &-6.8 & 7.3 & 17.4 \\
UKS/BLYP   &-12.5    & 12.9 & 34\\
\hline\hline 
\end{tabular}
\label{table:biradicalMAE}
\end{threeparttable}
\end{table}

The performance of UKS/BLYP is unsurprisingly poor, given the high level of spin-contamination in the singlet states revealed in Table \ref{table:SpinContam}.  The notable onset of larger errors in Fig. \ref{fig:biradicals_largeCAS}, i.e. for NH, NF, O$_2$, OH$^+$, O, C, and Si, is found to roughly correlate with the presence of spin-contamination in the unrestricted wavefunctions.  The Yamaguchi correction clearly improves upon the UKS/BLYP results, however the MAE of 7.3 kcal/mol and MaxE of 17.4 are still very large.  W2X is relatively more robust, with an MAE of 3.1 kcal/mol; however, the MaxE of 6.9 kcal/mol illustrates the difficulty in describing biradical systems even with ``gold-standard'' single-reference methods.  

In contrast, the AFQMC/U approach shows a significant improvement in accuracy (all ph-AFQMC results using UHF, UB3LYP, and RHF trials are shown in the Supporting Information).  We note that the MaxE for AFQMC/UHF is for the H$_2$CC molecule.  The UHF wavefunction for the triplet state, which should largely be of single-reference nature, still exhibits significant spin-contamination, as seen in Table \ref{table:SpinContam}.  The Slater determinant derived from UB3LYP orbitals appears to be relatively uncontaminated, with an $\langle S^2 \rangle$ value of 2.03 while still benefiting from the additional variational freedom due to the use of unrestricted orbitals. When this is used as the trial wavefunction for ph-AFQMC, the resulting ST gap in the CBS limit is -49.9(9) kcal/mol, which significantly reduces the deviation from experiment from 3.1(7) to -1.3(9) kcal/mol.  While more data points are needed to validate a more general claim, this case suggests that when a single-reference spin-state exhibits spin-contamination, AFQMC/UB3LYP can improve the accuracy of ST predictions over AFQMC/UHF.  This is reflected in the protocol specified earlier for AFQMC/U, which utilizes UHF trial wavefunctions for triplet states when $1.9 \leq \langle S^2\rangle \leq 2.1$, and UB3LYP otherwise.

\begin{figure}[H]
    \centering
    \includegraphics[width=14cm]{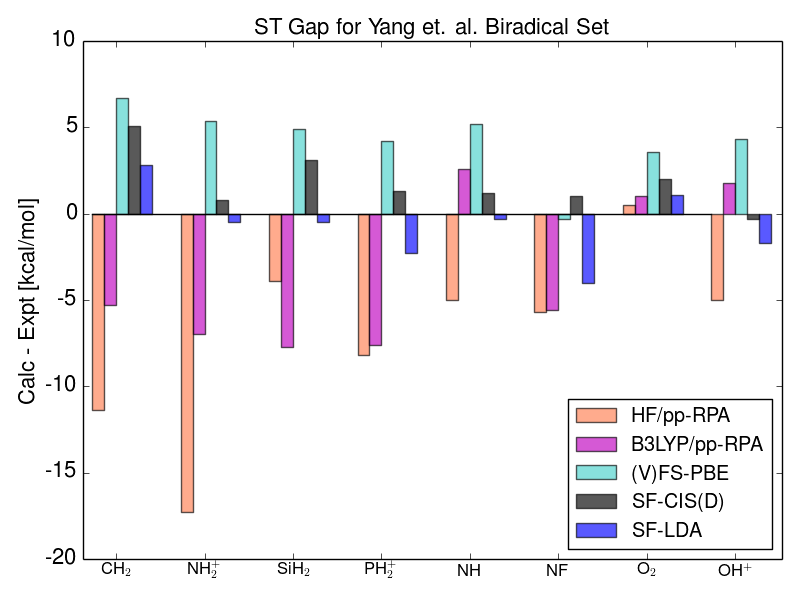} 
    \caption{Deviations [kcal/mol] of computational methods selected from Ref. \citenum{yang2015singlet} from experimentally-derived reference values, for a subset of 8 biradicals.} 
    \label{fig:YangComparison}
\end{figure}

To make a broader comparison regarding the ability of a variety of electronic structure methods (with similar computational scaling with respect to system size) to predict ST gaps in biradicals, we include calculated values from methods highlighted in Ref. \citenum{yang2015singlet} for a subset of 8 biradicals.  We plot this data in Fig. \ref{fig:YangComparison}, and provide a comparative statistical summary in Table \ref{table:biradicalYangMAE}.

\begin{table}[H]
\begin{threeparttable}
\caption{Mean signed, absolute, and maximum errors [kcal/mol] of ph-AFQMC results and other methods for the 8 molecule biradical subset shown in Fig. \ref{fig:YangComparison}.  Sorted by MAE value. Parenthesis indicate statistical errors.}
\begin{tabular}{l r r r}
\hline\hline 
           &  MSE  & MAE & MaxE    \\ [0.5ex] 
\hline 
AFQMC/U\tnote{$\dagger$}    & 0.83(7) & 1.0(7)  & 2.1(7)  \\
AFQMC/CAS\tnote{$\dagger$}    & 0.5(4)  & 1.1(4)  & 1.7(3)  \\
SF-LDA\tnote{a}        & -0.7 & 1.7  & 4.0  \\
SF-CIS(D)\tnote{b}  & 1.8   & 1.9  & 5.1 \\
W2X\tnote{c}        & 3.0  & 3.7  & 6.9  \\
(V)FS-PBE\tnote{d}  & 4.2  & 4.3  & 6.7  \\
B3LYP/pp-RPA\tnote{e}   & -3.5 & 4.8  & 7.7  \\
WA-KS/BLYP\tnote{c} & -6.6 & 6.6  & 17.5 \\
HF/pp-RPA\tnote{e}      &-7.0  & 7.1  & 17.3 \\ 
UKS/BLYP\tnote{c}   &-12.5 & 12.5 & 34.0 \\
\hline\hline
\end{tabular}
\begin{tablenotes}
\item[$\dagger$] This work.
\item[a] SF-TDDFT with LDA functional and noncollinear kernel, TZ2P basis, from Ref. \citenum{wang2005performance}.
\item[b] cc-pVQZ basis for NH, OH$^+$, NF, O$_2$; TZ2P basis for CH$_2$, NH$_2^+$, SiH$_2$, PH$_2^+$; from Ref. \citenum{KrylovSpinFlip}
\item[c] from Ref. \citenum{bao2016correlated}.
\item[d] (Variational) Fractional-Spin method, 6-311++G(2d,2p) basis, from Ref. \citenum{VariationalFSDFT}.
\item[e] aug-cc-pVDZ basis, from Ref. \citenum{yang2015singlet}.
\end{tablenotes}
\label{table:biradicalYangMAE}
\end{threeparttable}
\end{table}

For this subset of cases, there is no distinction between the AFQMC/UHF and AFQMC/U procedures, and AFQMC/U and AFQMC/CAS yield equivalent accuracy, considering statistical error bars.  Both produce MAEs of $\sim$1 kcal/mol and MaxEs of $\sim$2 kcal/mol, comparing favorably to all other methods.  As shown in the Supporting Information, the accuracy of AFQMC/UB3LYP and AFQMC/RHF is similar to that obtained via spin-flip methods for these systems.

%%%%%%%%%% benzynes %%%%%%%%%%%%
\subsection{Benzyne Isomers}

\begin{figure}[H]
    \centering
    \includegraphics[width=12cm]{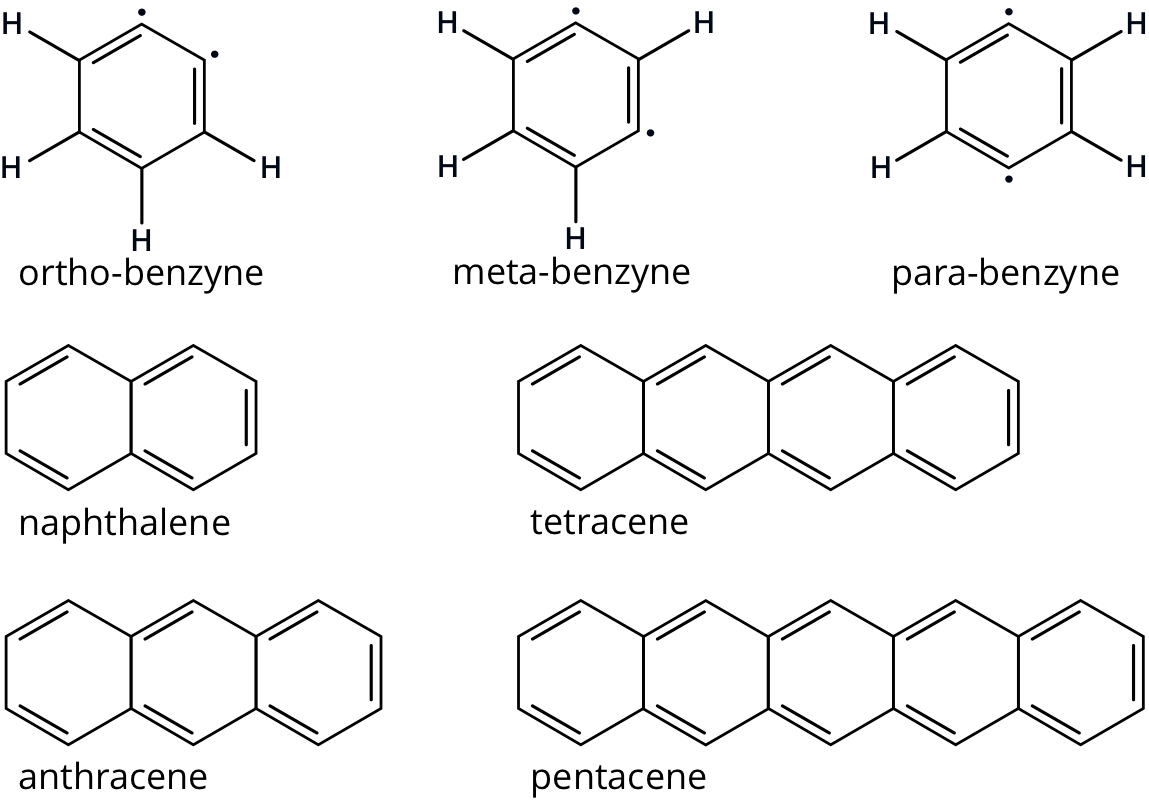}
    \caption{Benzyne isomers and polyacenes studied in this work.} 
    \label{fig:Structures}
\end{figure}

In this section we consider the ortho-, meta-, and para-benzyne isomers, shown in Fig. \ref{fig:Structures}, and compare predicted ST gaps with precise, gas-phase experimental measurements.  The ground state for all isomers is a singlet, and biradical character correlates with the distance between the unpaired electrons (ortho $<$ meta $<$ para).\cite{KrylovSpinFlip}.  These systems are of scientific interest in their own right, e.g. singlet  para-benzyne is a biradical that can abstract hydrogen atoms from specific positions in DNA, potentially enabling antitumor activity.\cite{sander1999m,crawford2001problematic}. 

\emph{Ab initio} calculations are difficult due to the strongly correlated biradical electrons.  In particular, singlet para-benzyne exhibits orbital instabilities at the RHF level, and subsequent RHF-based correlation methods produce poor results.  Better results are obtained with the broken-symmetry UHF reference, with complete spin symmetry-restoration via, e.g., subsequent CCSD calculation.\cite{crawford2001problematic}  With this in mind, we explore UHF and UB3LYP trial wavefunctions for ph-AFQMC, in addition to multi-determinant trial wavefunctions from CASSCF.  

Our ph-AFQMC calculations utilize geometries obtained from SF-DFT/6-311G* using the so-called 50/50 functional\cite{KrylovSpinFlip}.  The ortho- and meta- geometries are provided in Ref. \citenum{yang2015singlet}, the para- geometry in Ref. \citenum{bernard2012general}.  In Ref. \citenum{yang2015singlet}, ST gaps of para-benzyne are based on SF-CCSD/cc-pVTZ optimized geometries\cite{bernard2012general}.  However we have performed ph-AFQMC calculations on this SF-CCSD geometry with CASSCF trial wavefunctions utilizing increasingly large active spaces (up to 8 electrons in 16 orbitals), and still find a residual error from experiment of 2.9(7) kcal/mol in the cc-pVTZ basis (which is altered by only 1 kcal/mol in the estimated CBS limit).  For this reason, and for consistency with the ortho- and meta- isomers, we use the SF-DFT/6-311G* geometry for the para-isomer as well.

It is noteworthy that Refs. \citenum{yang2015singlet} and \citenum{SFTDDFTmhg} do not account for ZPE contributions in comparing calculated electronic ST energies to experimentally measured quantities.  For meta-benzyne this correction is $\sim$1 kcal/mol, so we choose here to subtract out ZPE values, taken from Ref. \citenum{KrylovSpinFlip}, from the experimental data when comparing with purely electronic predictions. 

$\langle S^2 \rangle$ values of the candidate unrestricted trial wavefunctions are shown in Table \ref{table:SpinContamBenzynes} for the benzyne isomers.  The UHF singlet and triplet states are both significantly contaminated in all isomers, with the singlet state of para-benzyne is severe case ($S^2$=1.68).  UB3LYP reduces the amount of spin-contamination in all cases, however it does not always eliminate it, e.g. $\langle S^2 \rangle$ of singlet para-benzyne is reduced to 0.92.  The AFQMC/U method will use UB3LYP trial wavefunctions for all isomers, since $\langle S^2 \rangle_{singlet}^{UHF} > 1.1$ for ortho- and para- benzynes, and $\langle S^2 \rangle_{triplet}^{UHF} > 2.1$ for all.

\begin{table}[H]
\begin{threeparttable}
\caption{$\langle S^2 \rangle$ of the singlet and triplet unrestricted solutions for ortho- meta- and para- benzyne molecules.}
\begin{tabular}{l r r r r }
\hline\hline 
           &  UHF singlet & UHF triplet & UB3LYP singlet & UB3LYP triplet \\ [0.5ex] 
\hline 
ortho   &  1.26  &  2.31  & 0.00 & 2.01 \\
meta    &  0.97  &  2.68  & 0.10 & 2.02 \\
para    &  1.68  &  2.31  & 0.92 & 2.01 \\
\hline\hline 
\end{tabular}
\label{table:SpinContamBenzynes}
\end{threeparttable}
\end{table}

\begin{figure}[H]
    \centering
    \includegraphics[width=16cm]{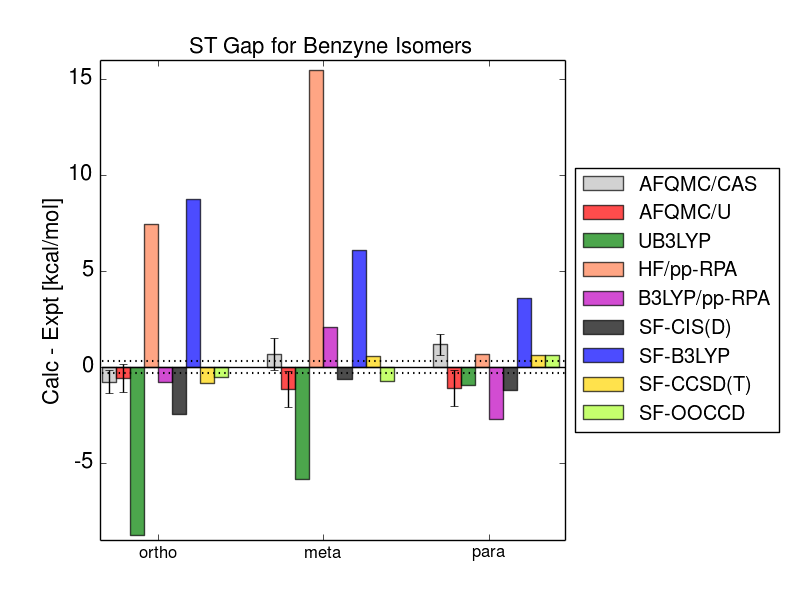} 
    \caption{Deviations [kcal/mol] of various calculations from ZPE-corrected experimental measurements.  Error bars show the statistical error of the QMC measurements.  Dotted black lines represent the reported uncertainty of the experimental measurements. } 
    \label{fig:benzyneFig}
\end{figure}

The resulting ST gaps are shown in Table \ref{table:benzyneST}, and the deviations of the predicted values from ZPE-corrected experimental results are shown in Fig. \ref{fig:benzyneFig}.  AFQMC/UHF and AFQMC/UB3LYP values are shown separately in the Supporting Information.

\begin{table}[H]
\begin{threeparttable}
\caption{ST gaps [kcal/mol] for the ortho- meta- and para- benzyne isomers. Parenthesis indicate statistical errors.}
\begin{tabular}{l r r r }
\hline\hline 
                & ortho & meta & para \\ [0.5ex]
\hline
expt\tnote{*}            &  37.5 $\pm$ 0.3 & 21.0 $\pm$ 0.3 & 3.8 $\pm$ 0.3 \\  
ZPE\tnote{**}     &  -0.6 & 1.0 & 0.5   \\ 
ZPE-corr'd expt &  38.1 & 20.0 & 3.3  \\
\hline
AFQMC/CAS\tnote{$\dagger$}         &  37.4(6) & 20.7(8) & 4.5(5) \\
AFQMC/U\tnote{$\dagger$}      &  37.6(7) & 18.9(9) & 2.2(9) \\
UB3LYP\tnote{a}          &  29.4    & 14.2    & 2.4    \\
HF/pp-RPA\tnote{b}          &  45.6    & 35.5    & 4.0    \\
B3LYP/pp-RPA\tnote{b}       &  37.4    & 22.1    & 0.6    \\
SF-CIS(D)\tnote{c}       &  35.7    & 19.4    & 2.1    \\
SF-B3LYP\tnote{d}        &  46.9    & 26.1    & 6.9    \\
SF-CCSD(T)\tnote{e}      &  37.3    & 20.6    & 4.0    \\
SF-oo-CCD\tnote{c}       &  37.6    & 19.3    & 3.9    \\
\hline\hline
\end{tabular}
\label{table:benzyneST}
\begin{tablenotes}
\item[*] Ref. \citenum{wenthold1998ultraviolet}
\item[**] SF-DFT/6-311G*, Ref. \citenum{KrylovSpinFlip}.
\item[$\dagger$] SF-DFT/6-311G* geometries from Ref. \citenum{KrylovSpinFlip}, cc-pVTZ basis. This work.
\item[a] 6-31G* basis, Ref. \citenum{SFTDDFTmhg}
\item[b] Ref. \citenum{yang2015singlet}. pp-RPA calculations in aug-cc-pVDZ basis. o- and m- geometries from SF-DFT with the 50/50 functional (Ref.\citenum{SFTDDFTmhg}).  p- geometry from SF-CCSD, Ref. \citenum{bernard2012general}. 
\item[c] SF-DFT/6-311G* geometries, cc-pVTZ basis, Ref. \citenum{KrylovSpinFlip}.
\item[d] cc-pVTZ basis, Ref. \citenum{bernard2012general}.
\item[e] SF-DFT/6-311G* geometries, cc-pVTZ basis, Ref. \citenum{manohar2008noniterative}. 
\end{tablenotes}
\end{threeparttable}
\end{table}

SF-oo-CCD and SF-CCSD(T) both consistently obtain sub kcal/mol accuracy, however they are computationally infeasible for larger systems due to the respective $O(M^6)$ and $O(M^7)$ scaling.  We thus focus presently on methods with lower scaling. 

AFQMC/CAS and AFQMC/U produce predictions of comparable accuracy, with maximum errors just outside 1 kcal/mol.  AFQMC/UHF achieves good accuracy for the ortho- and meta- isomers, however the severely spin-contaminated singlet state of para-benzyne results in a relatively large overestimation of the ST gap.  For this latter system AFQMC/UB3LYP substantially reduces the deviation from experiment from 5.7(8) to 1(1) kcal/mol, and we note that similarly pronounced corrections are observed in pp-RPA calculations when the B3LYP reference for the ($N-2$)-electron system is used instead of HF.\cite{yang2015singlet}  AFQMC/U and AFQMC/UB3LYP are equivelent for this set of molecules, and thus we again find that AFQMC/U, utilizing only unrestricted single-determinant trials, can produce results of comparable accuracy to both AFQMC/CAS and experiment.

We must emphasize the importance of explicitly breaking the spin-symmetry when obtaining UHF and UB3LYP trial wavefunctions, and choosing the solution with lowest-energy.  For instance, the calculated ST gap using a spin-pure ($\langle S^2 \rangle$ = 0) B3LYP trial for para-benzyne is -15 kcal/mol, vs +2.2 kcal/mol when the lower-energy $\langle S^2 \rangle =0.92$ unrestricted singlet state is used.

As a final remark in this section, we note that while indeed the errors from UB3LYP (i.e. without subsequent QMC) are generally reduced in comparison with those from the small molecule biradical set, we still find significant errors of -8.7, -5.8, and -0.9 kcal/mol for the ortho-, meta-, and para-benzynes.  UB3LYP systematically underestimates the ST gaps, due to the unrealistically low energy of the broken-symmetry singlet state.  Rather unexpectedly, however, the magnitude of the errors here are inversely correlated with diradical character.

These results suggest that for the benzyne isomers, which exhibit strong biradical character while sharing features similar to the planar aromatic ring systems relevent to chemical photocatalysis, the ST gaps are accurately predicted by ph-AFQMC with both CASSCF and single-determinant trials, and in the cc-pVTZ basis.  The accuracy of AFQMC/UHF, even in its spin projected form, is compromised by the heavily spin-contaminated singlet state in para-benzyne, though AFQMC/UB3LYP provides an improved prediction, and is utilized in our AFQMC/U method.

%%%%%%%%%% acenes %%%%%%%%%%%%
\subsection{Polyacenes}

Having shown that AFQMC/U can accurately describe molecules with strong biradical nature, we now show that this computational approach can scale to larger molecules, focusing on polyacenes from naphthalene to pentacene, shown in Fig. \ref{fig:Structures}.  These molecules all have singlet ground states.  We use geometries from Ref. \citenum{ibeji2015singlet} for the acenes, which were computed at the unrestricted B3LYP/6-31G(d) level of theory.  

Table \ref{table:SpinContamAcenes} shows that the extent of spin-contamination in the singlet UHF states increases with the number of fused rings. In contrast, wavefunctions constructed from UB3LYP orbitals are spin-pure, consistent with previous computational studies\cite{zhang2007lowest,FPAacenes, FPAacenesFirst}.

\begin{table}[H]
\begin{threeparttable}
\caption{$\langle S^2 \rangle$ of the singlet and triplet unrestricted solutions for polyacenes $n=2-5$.}
\begin{tabular}{l r r r r }
\hline\hline 
$n$           &  UHF singlet & UHF triplet & UB3LYP singlet & UB3LYP triplet \\ [0.5ex] 
\hline 
2    &  1.10     & 2.30   & 0 & 2.02  \\
3    &  1.78     & 2.68   & 0 & 2.02 \\
4    &  2.43     & 2.91   & 0 & 2.03 \\
5    &  3.06     & 3.44   & 0 & 2.03 \\
\hline\hline 
\end{tabular}
\label{table:SpinContamAcenes}
\end{threeparttable}
\end{table}

ST gaps calculated with ph-AFQMC in the cc-pVTZ basis are shown in Table \ref{table:aceneST}, alongside predictions from state-of-the-art \emph{ab initio} methods and available experimental data.  When more than one experimental value is given in Ref. \citenum{Evangelista_sCI}, we choose the value that is closest to that shown in Ref. \citenum{YangPNAS}.  Zero point energy corrections, which are subtracted out of the experimental values, are required in order to compare calculated electronic energies with experiment, and we utilize numbers from Ref. \citenum{Evangelista_sCI} derived from B3LYP/6-31G(d) geometry optimizations and frequency calculations. 

It is important to recognize that all the experimental values, except in the case of anthracene, \emph{cannot} be fairly compared directly with gas-phase calculations.   Ref. \citenum{FPAacenes} conveniently provides details of many of the experimental measurements, which are reproduced here.  While the adiabatic ST gap of anthracene was obtained via gas-phase photoelectron spectroscopy, the reported experimental measurement for naphthalene was done in ether-isopentane-alcohol (solid) solvents at 77K.  The tetracene measurement was done in poly(methyl methacrylate) matrix at 298K, and pentacene was measured in a tetracene matrix at 298K.  Clearly, the gas-phase 0K conditions assumed in our calculations are not consistent with the realistic experimental conditions for most of the polyacenes studied here.

\begin{table}[H]
\begin{threeparttable}
\caption{ST gaps [kcal/mol] for the polyacenes $n$=2-5.  Square brackets indicate alternate experiments, and parenthesis indicate statistical errors.}
\begin{tabular}{l r r r r }
\hline\hline 
                & naphthalene & anthracene & tetracene & pentacene \\ [0.5ex]
\hline
expt\tnote{*}   &  [60.9] 61.0  & [42.6] 43.1 & 29.4 & 19.8$\pm$0.7 \\  
ZPE\tnote{*}    & -3.4  & -2.3 & -1.8 & -1.5 \\ 
ZPE-corr'd expt &  64.4 & 45.4 & 31.2 & 21.3 \\
\hline
AFQMC/U\tnote{a}      &  68.0(1.2) & 46.2(1.2) & 34.0(1.6) & 25.2(1.6) \\
UB3LYP\tnote{b}          &  62.6 & 41.8 & 27.7 & 17.9 \\
CCSD(T)/FPA\tnote{c}     &  65.8 & 48.2 & 33.5 & 25.3 \\
B3LYP/pp-RPA\tnote{d}        & 66.2 & 45.7 & 32.1 & 22.6  \\
GAS-pDFT (FP-1)\tnote{e} & 70.6 & 45.5 & 33.6 & 25.4 \\
GAS-pDFT (WFP-3)\tnote{e}& 64.7 & 43.1 & 28.8 & 20.5  \\
ACI-DSRG-MRPT2\tnote{f}  & 62.2 & 43.2 & 28.3 & 18.0  \\
DMRG-pDFT\tnote{g}       & 67.1 & 46.1 & 31.6 & 22.6  \\
\hline\hline
\end{tabular}
\label{table:aceneST}
\begin{tablenotes}
\item[*] Taken from Ref. \citenum{Evangelista_sCI}
\item[a] this work
\item[b] UB3LYP/6-31G(d) geometries and energies. Ref. \citenum{hachmann2007radical}
\item[c] B3LYP/cc-pVTZ geometries. Ref. \citenum{FPAacenes}
\item[d] UB3LYP/6-31G* geometries, B3LYP reference, cc-pVDZ basis. Ref. \citenum{YangPNAS}
\item[e] UB3LYP/6-31G(d,p) geometries, tPBE/6-31G(p,d). Active spaces defined in Ref. \citenum{GhoshST}
\item[f] UB3LYP/6-31G(d) geometries. Ref. \citenum{Evangelista_sCI}
\item[g] UB3LYP/6-31G(d,p) geometries. tPBE/6-31+G(p,d). Ref. \citenum{SharmaAcene}
\end{tablenotes}
\end{threeparttable}
\end{table}

\begin{comment}
\begin{figure}[H]
    \centering
    \includegraphics[width=14cm]{./Figures/acenes_UQMC.png} 
    \caption{Deviations [kcal/mol] of various calculations from ZPE-corrected experimental measurements.  Error bars show the statistical error of the QMC measurements.  } 
    \label{fig:acenes}
\end{figure}
\end{comment}

We observe that the ph-AFQMC predictions are insensitive to the trial wavefunction used for these polyacene systems, and show results from UHF, RHF, and UB3LYP trial wavefunctions in the Supporting Information.  Given that naphthalene through pentacene exhibit little biradical character\cite{YangPNAS}, the extreme spin-contamination of the UHF solutions shown in Table \ref{table:SpinContamAcenes} is likely \emph{not} representative of strong electron correlation, and hence restricted trial wavefunctions can also be expected to yield accurate results.  Moreover, all ph-AFQMC variants unambiguously achieve high accuracy with respect to the gas-phase measurement for anthracene.  The ph-AFQMC predictions for the other polyacenes, which were experimentally probed in (solid) solvent matrix, appear to systematically overestimate the experimental values by a few kcal/mol.   For naphthalene, we performed ph-AFQMC calculations with CASSCF(10e,10o) trial wavefunctions, which gave a ST gap of 67(1) kcal/mol.  This value is in agreement with that from AFQMC/U, given statistical error bars, and lies above the reported ZPE-corrected experiment by some 3 kcal/mol.  Very similar overestimations by AFQMC/U are seen for $n$=2,4,5 and are corroborated by CCSD(T)/FPA, B3LYP/pp-RPA, and DMRG-pDFT (though not by UB3LYP and ACI-DSRG-MRPT2).  Thus it seems reasonable to hypothesize that the calculations' neglect of molecular environment may be responsible, though admittedly there are a number of other factors that might be expected to contribute.  For instance, adiabatic ST gaps are known to be sensitive to the optimized geometries, which can result in variations on the order of 1-3 kcal/mol.\cite{Evangelista_sCI}   

Overall, given the uncertainties due to the treatment of temperature, solvent, and molecular geometries in these acene calculations, ph-AFQMC with single-determinant trial wavefunctions gives satisfactory agreement with both experiments and other highly-accurate electronic structure methods, and moreover can scale with near-perfect parallel efficiency to systems as large as pentacene in a triple-zeta basis, which has 146 electrons and 856 basis functions.  

%%%%%%%%%%%%%%%%%%%%
\section{Conclusions}
%%%%%%%%%%%%%%%%%%%%

With CASSCF trial wavefunctions, ph-AFQMC can predict ST gaps with sub kcal/mol accuracy with respect to gas-phase experimental measurements for a set of 13 small molecules with singlet states of strong biradical character.  However, for large systems the generation of such a trial wavefunction quickly becomes impractical.  The main result of this work is that near-chemical accuracy for gas-phase ST gaps can also be obtained, even for strongly correlated biradical systems, with a spin-projected ph-AFQMC technique, which initializes walkers with restricted determinants while using an unrestricted single-determinant trial wavefunction to implement the phaseless constraint.  We establish a quantitative criteria for choosing UHF or UB3LYP orbitals based on the spin-contamination of the UHF wavefunction, and the resulting AFQMC/U methodology is validated on the small molecule test set, the ortho-, meta-, and para-benzyne isomers, and of all the polyacenes for which experimental results are reported (though a true gas-phase experiment is only available for anthracene).  The ph-AFQMC method is shown to provide a balanced and robust approach to accurately predicting ST gaps for molecules relevant to chemical photocatalysis.  

With computational cost scaling as $O(M^4)$ or lower,\cite{motta2018efficient} and with near-perfect parallel efficiency,\cite{shee2019,shee2018gpu} these calculations take only minutes of wall-time on GPU-accelerated supercomputing resources.  While further investigation will be required to realistically treat solvation effects, this study paves the way for future work that will focus on a large set of both known and potential photocatalysts.  Another possible extension we plan to undertake in the future is the computation of spin gaps in strongly correlated solids.

\begin{acknowledgement}
J.S. acknowledges Roel Tempelaar, Joonho Lee, Mario Motta, Hao Shi, Ian Dunn, Andrew Pun, and Zack Strater for insightful discussions. 
This research used resources of the Center for Functional Nanomaterials, which is a U.S. DOE Office of Science Facility, and the Scientific Data and Computing Center, a component of the Computational Science Initiative, at Brookhaven National Laboratory under Contract No. DE-SC0012704.
An award of computer time was provided by the INCITE program.  This research used resources of the Oak Ridge Leadership Computing Facility, which is a DOE Office of Science User Facility supported under Contract DE-AC05-00OR22725.  The Flatiron Institute is a division of the Simons Foundation.  D.R.R. acknowledges support from NSF CHE 1839464, and S.Z. from DOE DE-SC0001303.
% ACS style
\end{acknowledgement}

\begin{suppinfo}

%This will usually read something like: ``Experimental procedures and
%characterization data for all new compounds. The class will
%automatically add a sentence pointing to the information on-line:

\end{suppinfo}

\bibliography{achemso-demo}

\end{document}